\documentclass[onecolumn]{genj}
\pdfoutput=1
\usepackage{amsmath}
\usepackage{fleqn}
\usepackage{times}
\usepackage{graphicx}
\usepackage{color}
\def\bsym{\boldsymbol}

\begin{document}
\hyphenchar\font = -1
\sloppy
\mathindent=2mm
\eqsecnum

\title{Nonlinear elasticity under moderate to strong compression}

\author[B.L.N. Kennett]{B.L.N.. Kennett\\
Research School of Earth Sciences, The Australian National University, Canberra, ACT 2601, Australia, \\ (Brian.Kennett@anu.edu.au) }

\maketitle

\begin{abstract}
The strain-energy formulation of nonlinear elasticity can be extended to the case of significant
compression by modulating suitable strain energy terms by a function of relative volume.
For isotropic materials this can be accomplished by the product of representations of shear, in 
terms of the invariants of the Seth-Hill family of strain measures, and a function of volume.
The incremental shear modulus under pressure is determined by this function, but nonlinear 
effects are retained for large strains.  Suitable functional forms can be derived from existing 
equations of state for moderate to strong compression.
For anisotropic materials, a similar development can be made directly with strain energy terms 
depending directly on the Seth-Hill strain tensors. Shear aspects can be emphasised by exploiting
the equivoluminal components of the strain tensors.
Such formulations may be helpful for materials under the conditions prevailing in the Earth's interior.
\end{abstract}


\section{Introduction}

Many applications of nonlinear elasticity are concerned with extensional environments with 
emphasis on shear properties, a useful review is provided by Mihai \& Goriely (2017).
Compressibility has commonly been neglected in the nonlinear case, but has been recognised 
to be significant in studies of soft tissues (e.g. Beex, 2019).
In contrast, in the study of the properties of materials at high pressures the emphasis has been on the 
development of equations of state for the bulk modulus.
Improved experimental and computational procedures mean that incremental shear properties from a 
compressed state have become accessible, and so a full constitutive equation is needed (Kennett, 2017).
Large shears tend to be suppressed as pressure increases, but can be significant in the Earth's 
lithosphere.

Many of the formulations of shear properties are based on the superposition of functions of
members of  the Seth-Hill strain tensors (Seth, 1964; Hill 1968) and their 
associated conjugate stresses, which 
provide extensions of Hooke's law. The members of this suite of strain measures are characterised 
by the exponent on the principal stretches.
We here show how standard nonlinear strain energy formulations can be adapted to carry
shear properties into the compressional regime, with the aid of an auxiliary function of density
modulating a deviatoric term.

For Earth materials, a semi-empirical linear relationship between the incremental shear modulus, 
the bulk modulus and pressure can be used to specify suitable functional forms for the 
auxiliary function.
By this means a shear modulus distribution can be associated with existing equations of state to 
provide a full constitutive equation..
This strain energy formulation is simplest in the isotropic case, but can be adapted to 
the anisotropic case by using the full strain tensors rather than their invariants.

\section{Isotropic materials under pressure}
\renewcommand\onethird{\textstyle{\frac{1}{3}}\displaystyle}
\renewcommand\onehalf{\textstyle{\frac{1}{2}}\displaystyle}

We consider a deformation from a reference state (unstressed) described by
coordinates $\bsym{\xi}$ to a current state described by coordinates
$\mathbf{x}$. The relation between the states is provided by the
deformation gradient tensor $\mathbf{F} = \partial \mathbf{x}/\partial \bsym{\xi}$,
and $ J = \det \mathbf{F} = V/V_0$ is then the ratio of a volume element
in the current state ($V$) to that in the reference state ($V_0$).
We introduce a strain energy $W(\mathbf{F})$ depending on deformation, which specifies the
constitutive equation for a material.

In terms of $\mathbf{F}$ and the Green strain 
$\mathbf{E} = \frac{1}{2}(\mathbf{F}^T \mathbf{F} - \mathbf{I})= \frac{1}{2}(\mathbf{C} - \mathbf{I})$,
the components of the stress tensor $\bsym{\sigma}$ are given by
\begin{equation}
J \sigma_{ij} = F_{ik} \frac{\partial W}{\partial F_{jk}} =
F_{ik} F_{jl} \frac{\partial W}{\partial E_{kl}} ,
\label{e.1}
\end{equation}
where we use the Einstein summation convention of summation over repeated
suffices.

The deformation gradient $\mathbf{F}$ can be written in terms of a stretching 
component and a rotation in two ways
\begin{equation}
\mathbf{F} = \mathbf{R}\mathbf{U} = \mathbf{V}\mathbf{R}
\label{e.2}
\end{equation}
where $\mathbf{U}^2 = \mathbf{F}^{T}\mathbf{F} = \mathbf{C}$ and 
$\mathbf{V}^2 = \mathbf{F}\mathbf{F}^{T} = \mathbf{B}$.
The matrices $\mathbf{U}$, $\mathbf{V}$ have the same eigenvalues, 
the principal stretches $\lambda_1, \lambda_2, \lambda_3$,
but the principal axes vary in orientation by the rotation $\mathbf{R}$.

The Seth-Hill class of strain measures take the form:
\begin{eqnarray}
\mathbf{E}_q (\mathbf{U}) = 
\begin{cases} 
\frac{1}{q} (\mathbf{U}^q - \mathbf{I})& \text{if}\ q \ne 0, \\
\ln{\mathbf{U}} & \text{if}\ q = 0,
\end{cases}
\label{eo.3}
\end{eqnarray}
where $\mathbf{I}$ is the identity tensor.
The Green strain is thus $\mathbf{E}_2$.  All the members of this class of strain measures take the
same form for infinitesimal deformation.

The separation between volumetric deformation and shear-type deformation, which is equivoluminal, 
can be achieved by working with $J$ and the normalised deformation gradient
$\mathbf{F}^{*} = J^{-1/3} \mathbf{F}$, so that $\det \mathbf{F}^{*} = 1$.

For an isotropic medium, the strain energy $W$ can be represented as a 
function of invariants of strain measures (e.g. Spencer, 1980).
Useful invariants of $\mathbf{U}$, $\mathbf{V}$ are
\begin{equation}
J = \lambda_1 \lambda_2 \lambda_3 =  \det \mathbf{U} ,
\label{eo.4}
\end{equation}
a purely hydrostatic term, representing changes in volume,
and the set
\begin{equation}
L_q= J^{-q/3} [ \lambda_1^q  + \lambda_2^q  + \lambda_3^q] 
=  J^{-q/3} \mathrm{tr}\{\mathbf{U}^q\},
\quad q \ne 0,
\label{e.5}
\end{equation}
which concentrate on the deviatoric aspects of deformation.
Note that $\frac{1}{q}\{ L_q- 3 \}$ corresponds to the trace of the equivoluminal
part of the Seith-Hill tensors, evaluated in terms of $\mathbf{U}^{*} = J^{-1/3} \mathbf{U}$.

For an isotropic medium the principal axes of the stress tensor $\bsym{\sigma}$
align with those of $\mathbf{V}$, $\mathbf{B}$ (the Eulerian triad), whereas the principal axes of
$\mathbf{U}$, $\mathbf{C}$ and $\mathbf{E}$ are rotated by $\mathbf{R}$ (the Lagrangian triad).
In terms of the principal stretches we can recast (\ref{e.1}) in the
form of an expression for the $r$th principal stress 
\begin{equation}
\sigma_r = \frac{1}{J} \lambda_r \frac{\partial {W}}{\partial \lambda_r} ,
\qquad \textrm{no sum on } r,
\label{e.6}
\end{equation}
whilst recognising the rotation between the principal directions of the elements
on the left- and right-hand sides of the equation (\ref{e.6}). 

Many of the formulations for nonlinear shear given by Mihai \& Goreiley (2017) can be 
expressed as a linear combinations
of the $L_q$ invariants, with constant coefficients. Under compression 
Kennett (2017) has shown that it is possible to associate a shear component to
existing equations of state, linking pressure and volume, by introducing a deviatoric term 
modulated by a function of  volume into the strain energy.
The specific form used in Kennett (2017) was derived from that for a 
neo-Hookean solid in terms of $L_2$,
but can be generalised to allow for a more complex shear behaviour.

Consider a strain energy function ${W}$ as a function of stretch
invariants $J$, $\{L_q\}$ with two independent volume terms $\Phi(J)$ and $\Psi(J)$:
\begin{equation}
{W} = \Phi(J) + \Psi(J) \sum_q a_q \frac{1}{q}\{ L_q- 3 \}, \quad \textrm{with}  \ \ \sum_q a_q =1,
\label{e.7}
\end{equation}
incorporating a direct volume dependence in $\Phi(J)$ and a deviatoric
component in the second term.  As noted above this is equivalent to an expansion in
terms of equivoluminal Seth-Hill tensors.
\newline
For \textit{purely hydrostatic compression}: 
$\lambda_1 = \lambda_2 = \lambda_3 = \hat{\lambda}$,
$J = \hat{\lambda}^3$ 
and $\sum_q a_q\frac{1}{q} \{L_q-3\} = \sum_q a_q \frac{1}{q}\{\hat{\lambda}^{-q}3\hat{\lambda}^q-3\} = 0$, \newline
so that the deviatoric term $\sum_q a_q\frac{1}{q} \{L_q - 3\}\, \Psi(J) = 0$.

For the strain energy  (\ref{e.7}) with both compressional and deviatoric components,
the $r$th principal stress takes the form:
\begin{equation}
\sigma_r = 
\frac{\partial\Phi}{\partial J} + \frac{\partial\Psi}{\partial J} \sum_q a_q \frac{1}{q} \{ L_q- 3 \}
+ \frac{1}{J} \Psi(J) \sum_q a_q  J^{-q/3}\{ \lambda_r^q-\onethird [ \lambda_1^q  + \lambda_2^q  + \lambda_3^q ]\} .
\label{e.8}
\end{equation}
The full stress tensor $\bsym{\sigma}$ can therefore be written as
\begin{equation}
\bsym{\sigma}  = \mathbf{R} \left\{
\left[\frac{\partial\Phi}{\partial J} + \frac{\partial\Psi}{\partial J} \sum_q a_q \frac{1}{q} \{ L_q- 3 \}  \right]\mathbf{I}
+
\frac{1}{J} \Psi(J) \sum_q a_q  J^{-q/3}\left\lbrace \mathbf{U}^q - \onethird  \mathrm{tr}\{\mathbf{U}^q\} 
\mathbf{I} \right\rbrace
\right\} \mathbf{R}^T .
\label{e.9}
\end{equation}
For the purely hydrostatic case, the deviatoric terms vanish and the stress tensor reduces to
\begin{equation}
- p \mathbf{I}  = \frac{\partial\Phi}{\partial J} \mathbf{I} .
\label{e.10}
\end{equation}

The incremental elastic moduli about this hydrostatically compressed state can be extracted
from the stress tensor (\ref{e.9}) by making a first order expansion with 
$\lambda_r = \hat{\lambda}(1 + e_r)$, so  that 
$J = \hat{\lambda}^3 [1 +\mathrm{tr}\{\mathbf{e}\}] + O(e^2)$.
In this case the $r$th principal stress takes the form
\begin{equation}
\sigma_r = -p + J \frac{\partial^2 \Phi}{\partial J^2}\, \mathrm{tr}\{\mathbf{e}\} +
\frac{1}{J} \Psi(J) \left( \sum_q q a_q \right) [e_r - \onethird \mathrm{tr}\{\mathbf{e}\} ] .
\label{e.11}
\end{equation}
The representation of the principal stress in terms of the bulk modulus $K$
and shear modulus $G$ is 
\begin{equation}
\sigma_r = -p
+ K \mathrm{tr}\{\mathbf{e}\} 
+ 2G \left( e_r - \onethird \mathrm{tr}\{\mathbf{e}\} \right)  ,
\label{e.12}
\end{equation}
and thus we identify the incremental moduli as:
\begin{equation}
K = J \frac{\partial^2 \Phi(J)}{\partial J^2},
\qquad
G = \frac{1}{2J} \Psi(J) \sum_q q a_q .
\label{e.13}
\end{equation}

The shear properties for incremental strain are thus determined by $\Phi(J)$, but for finite
strain will be modulated by the nature of the sum over the stretch invariants.
Thus allows a wide variety of behaviour to be captured.
The choice of the functions of volume $\Phi(J)$ and $\Psi(J)$ depends on the 
desired properties under pressure.  

A more general representation of incremental properties about an initial stress state 
in terms of the stretches $\{\lambda_i$\} has been provided by Destrade \& Ogden (2011), 
This treatment allows the possibility of non-hydrostatic scenario, but reduces to 
(\ref{e.12}) for a state of pure compression.

\begin{figure}
\centering
\resizebox{125mm}{!}
{\includegraphics*{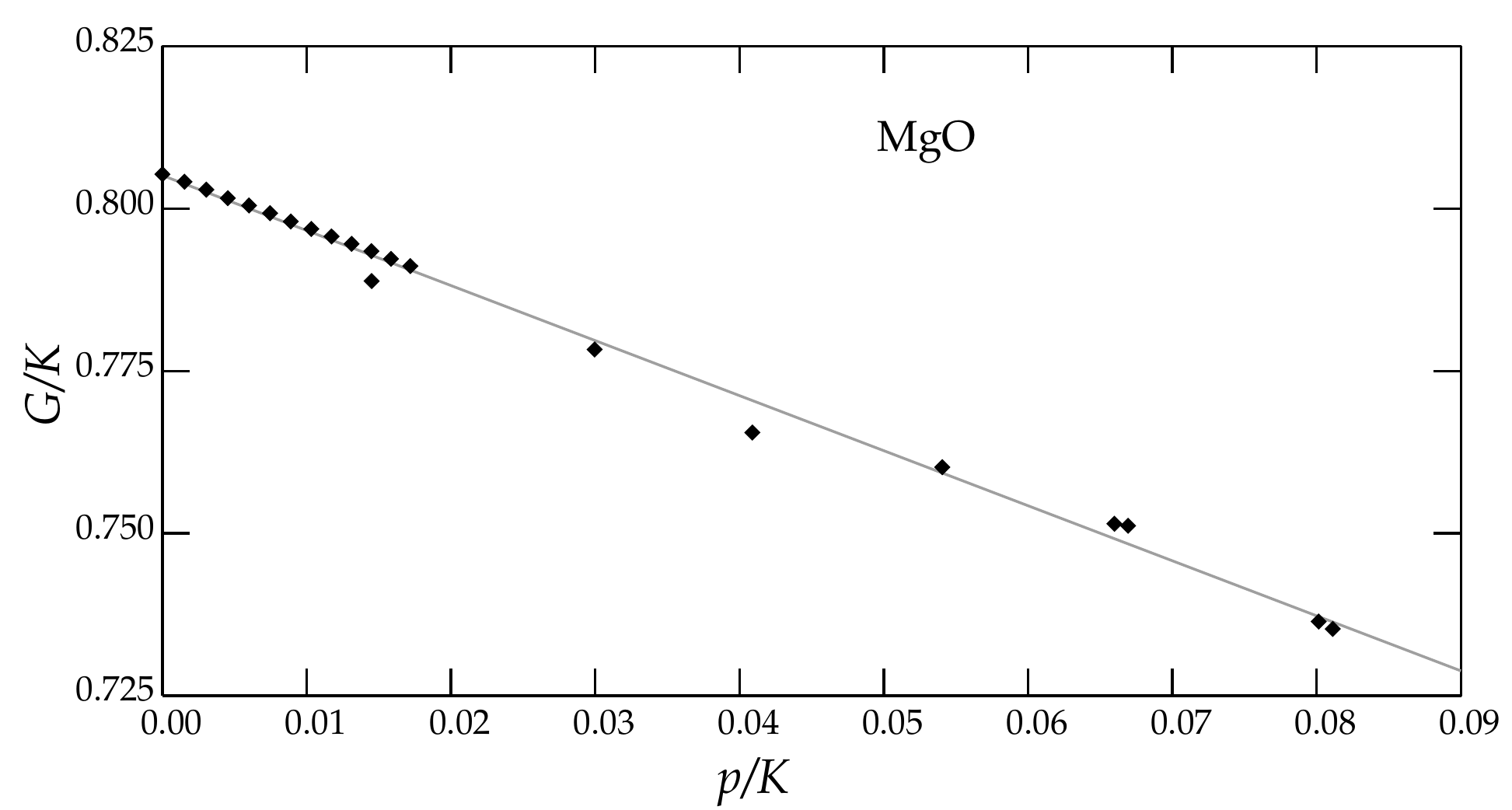}}
\caption{Illustration of the linear dependence of $G/K$ on $p/K$ to 20 GPa pressure, for the adiabatic 
bulk modulus $K$ of  periclase (MgO) using data from Jackson \& Niesler (1982), 
Sinogeikin \& Bass (2000), and Zha et al. (2000).}
\label{f.1}
\end{figure}

In applications to Earth materials, a number of different formulations have been developed 
for equations of state through  the strain energy term $\Phi(J)$ (see, e.g., Kennett, 2017).
Formulations for shear are much less common, and the most common form employed is
the  Birch-Murnaghan development in terms of powers of Eulerian strain 
(Stixrude \& Lithgow-Bertelloni, 2005).
The limitations of this approach for high pressures have been well documented by Stacey
\& Davis (2004), who advocate instead the Keane equation of state for bulk modulus, 
but this does not have an associated shear modulus. 
Kennett (2017) has shown how the semi-empirical relation
\begin{equation}
G = aK - bp,
\label{e.14}
\end{equation}
can be used to produce an effective representation of shear-properties under pressure.
Note that with the formulation above (.10,2.13) this means that $\Psi(J)$ is related to the derivatives of $\Phi(J)$, with  $\partial \Phi/\partial J$ from pressure $p$ and $\partial^2 \Phi/ \partial J^2$ from $K$.
In terms of the bulk and shear moduli at zero pressure ($K_0$, $G_0$) and their pressure 
derivatives ($K^{\prime}_0$, $G^{\prime}_0$)
\begin{equation}
a = \frac{G_0}{K_0}, \quad b = \Big(\frac{G_0}{K_0} \Big) K_0^{\prime} -G_0^{\prime}.
\label{e.15}
\end{equation}
Equations (\ref{e.14}, \ref{e.15}) provide a good representation of experimental results for minerals, 
as illustrated in Figure \ref{f.1} for the isotropic properties of MgO.

\begin{figure}
\centering
\resizebox{125mm}{!}
{\includegraphics*{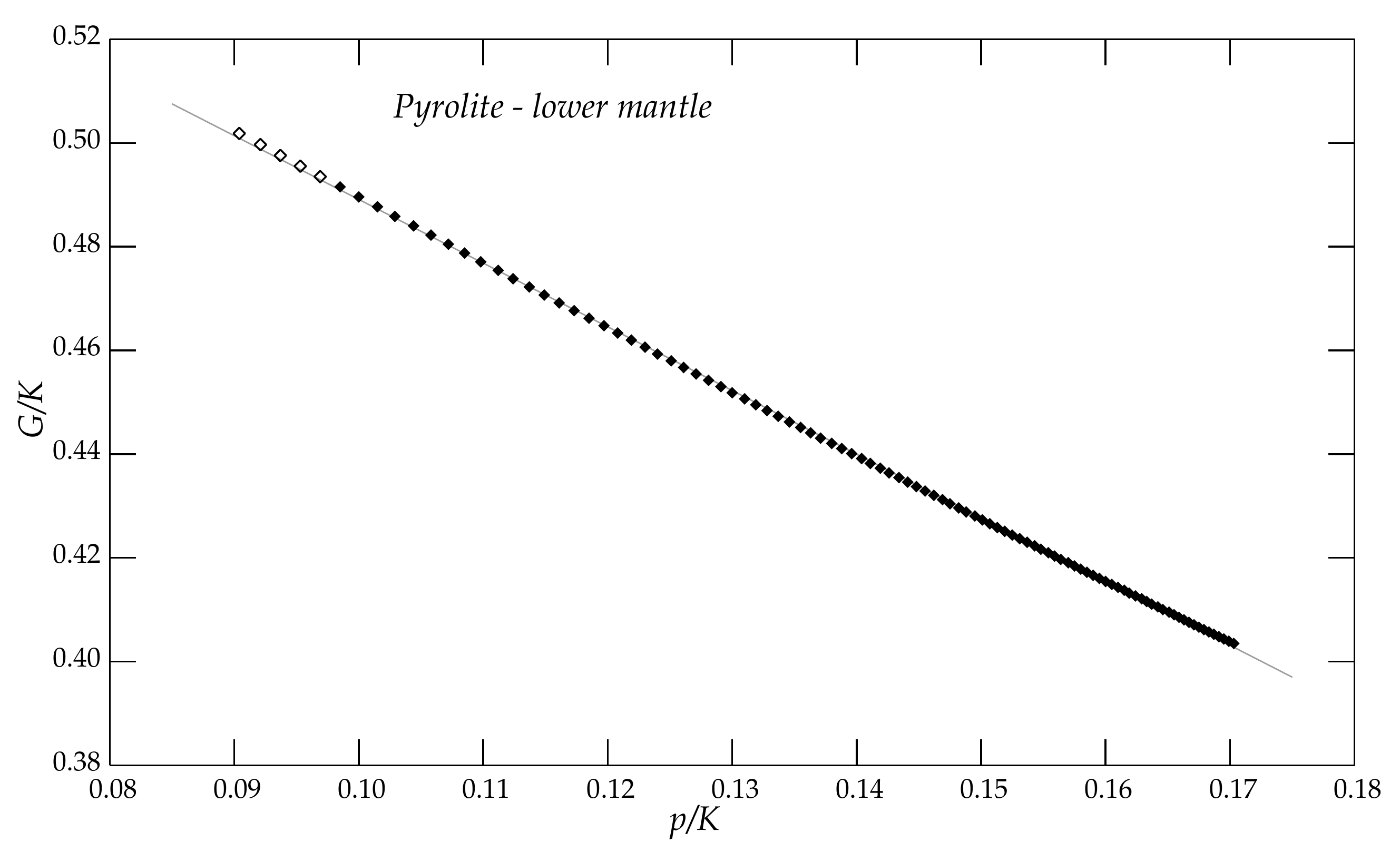}}
\caption{Illustration of the linear dependence of $G/K$ on $p/K$ for a pyrolite composition lower mantle mineral assemblage from Gr\'{e}aux et al. (2019). The open symbols indicate the zone where some residual garnet may be present.}
\label{f.2}
\end{figure}

The linear relation also provides a good description of the properties of mineral assemblages. 
We illustrate the results for the  Earth's lower mantle using the model developed by Gr\'{e}aux et al. 
(2019) in Figure \ref{f.2}.
The dominant minerals are bridgmanite and ferropericlase, and some residual majorite garnet is
present at the top of the lower mantle where there is a slight deviation from the linear trend.
Although the linear form works well over a large range of pressures (up to 140 GPa for the lower 
mantle), Burakovsky et al. (2004) suggest that (\ref{e.14}) should be modified with a slowly-varying
pressure dependence for $b$ to allow a match  to the expectation for infinite pressure.

The combination of the strain energy development (\ref{e.7}) with the identification of moduli 
(\ref{e.13}) and the relation (\ref{e.14}) provides a flexible way of extending nonlinear elastic 
effects to a compressed state, whilst retaining complex shear behaviour for finite strains.

\section{Anisotropic materials under pressure}

Many natural materials such as wood and tissues show distinct anisotropy in their properties. 
Most minerals have significant anisotropy, and its only in aggregate that much of the Earth
appears to have nearly isotropic properties. For the description of the behaviour of materials 
under strong compression it is therefore desirable to be able to provide a full description of the 
anisotropic behaviour.

In the general anisotropic situation the directions of the principal stretches do not remain constant 
and so their variation has to be taken into account in any formulation.
Even so it is possible to express the strain energy $W$ in terms of $J$ and the normalised 
deformation gradient $\mathbf{F}^{*}$ as $W(\mathbf{F}) = W^{*}(J,\mathbf{F}^{*})$.
With this representation the Cauchy stress tensor $\bsym{\sigma}$ is given by
\begin{equation}
\bsym{\sigma} = \frac{1}{J}\mathbf{F}\frac{\partial W}{\partial \mathbf{F}} =
\frac{\partial W^{*}}{\partial J} \mathbf{I} +
\frac{1}{J} \left[\mathbf{F}^{*}\frac{\partial W^{*}}{\partial \mathbf{F}^{*}} -  
\onethird  \mathrm{tr} \left\{\mathbf{F}^{*}\frac{\partial W^{*}}{\partial \mathbf{F}^{*}} \right\} 
\mathbf{I} \right] .
\label{e.21}
\end{equation}
Hence in asimilar way to the isotropic case above we can make a separation into pressure
dependence and shear deformations.

For orthotropic materials, Latorre \& Mont\'{a}ns (2017) have split the strain energy into an isotropic 
and a specifically anisotropic part.  Such an approach may well be suitable for small compressions, 
but when we want to include strong compression we should allow for volume dependence of 
all the components.
Building on the approach used for isotropy we look to combine a  volumetric 
component with equivoluminal term modulated by a function of relative volume.
We use the equivoluminal equivalents of the Seth-Hill measures (e.g.,  Miehe \& Lambrecht, 2001)
\begin{eqnarray}
{\mathbf{E}}^{*}_q (\mathbf{U}^{*}) = 
\begin{cases} 
\frac{1}{q} (\mathbf{U}^{*q} - \mathbf{I})& \text{if}\ q \ne 0, \\ 
\ln{\mathbf{U}^{*}} & \text{if}\ q = 0.
\end{cases}
\label{e.22}
\end{eqnarray}
and construct a strain energy function
\begin{equation}
W^{*} (J, \mathbf{F}^{*}) = \Phi(J) + \sum_q S_q(J) W_q (\mathbf{E}^{*}_q).
\label{e.23}
\end{equation}
Then in terms of the equivoluminal strain 
$\mathbf{E}^{*} = \mathbf{E}_2^{*} =\onehalf (\mathbf{U}^{*2} - \mathbf{I})$,
\begin{equation}
\mathbf{F}^{*}\frac{\partial W^{*}}{\partial \mathbf{F}^{*}} = 
\mathbf{F}^{*}\frac{\partial}{\partial \mathbf{E}^{*}}\sum_q S_q(J) W_q (\mathbf{E}^{*}_q)
\mathbf{F}^{*T}
= \mathbf{F}^{*} \sum_q \frac{\partial S_q(J)}{\partial \mathbf{F}^{*}}
W_q (\mathbf{E}^{*}_q) + \mathbf{F}^{*} \sum_q S_q(J) \frac{\partial W_q (\mathbf{E}^{*}_q)}
{\partial \mathbf{E}^{*}} .
\label{e.24}
\end{equation}
The derivative of the functions of relative volume 
\begin{equation}
\frac{\partial S_q(J)}{\partial\mathbf{F}^{*}} = 
\frac{\partial S_q(J)}{\partial J} \frac{\partial J}{\partial \mathbf{F}^{*}} = 
J \frac{\partial S_q(J)}{\partial J}  \mathbf{F}^{*-T} .
\label{e.25}
\end{equation}
Thus, the shear component of the Cauchy stress expression
\begin{equation}
\mathbf{F}^{*}\frac{\partial W^{*}}{\partial \mathbf{F}^{*}} = 
J \mathbf{I} \sum_q \frac{\partial S_q(J)}{\partial J} W_q (\mathbf{E}^{*}_q)
+\mathbf{F}^{*}
\sum_q S_q(J) \frac{\partial W_q (\mathbf{E}^{*}_q)}{\partial \mathbf{E}_q^{*}}
\bsym{\mathsf{P}}_q .
\label{e.26}
\end{equation}
The fourth order projection tensor $\bsym{\mathsf{P}}_q =
\partial\mathbf{E}^{*}_q/\partial \mathbf{E}^{*} $ is detailed in the Appendix.

The contributions to the stress (\ref{e.21}) from the functions $\{S_q(J)\}$ are purely hydrostatic,
and the shear dependence comes from the choices made for $\{W_q(\mathbf{E}^{*})\}$.
The evolution of the stress tensor under pressure and the consequent 
elastic properties can be evaluated by a perturbation treatment 
around a state of pure compression as in Section 2.
As before the bulk modulus $K$ is given by $J\partial^2 \Phi / \partial J^2$.

A simple form for the individual strain energy terms is quadratic:
\begin{equation}
W_q = \onehalf \mathbf{E}_q : \bsym{\mathsf{Z}}_q : \mathbf{E}_q ,
\label{e.27}
\end{equation}
where $\bsym{\mathsf{Z}}_q$ is a fourth-order stiffness tensor with 21 independent components,
and $:$ denotes the double inner product, so that $\mathbf{A} : \mathbf{B} = A_{ij}B_{ji}$.
With a sum of a number of Seth-Hill contributions a variety of deformation styles 
can be produced (e.g., Beex, 2019). 
In this case for hydrostatic stress  $\mathbf{E}^{*}_q$ vanishes, and as in the treatment of 
isotropic elasticity in Section 2 a perturbation treatment about a hydrostatic state simplifies 
significantly to leave a shear contribution specified by the $S_q(J)$.

In this anisotropic development we have introduced separate functions of relative volume 
for each order of the Seth-Hill strain measures, but simplified forms may be preferable.
If the anisotropic properties are consistent with increasing pressure, 
a suitable strain energy formulation for a material under moderate 
compression would be
\begin{equation}
W = \Phi(J) + S_A(J) \left[  \mathbf{E}^{*}_{-1} : \bsym{\mathsf{A}}: \mathbf{E}^{*}_{-1}  \right],
\label{e.28}
\end{equation}
in terms of the equivoluminal component of the Almansi strain 
${\mathbf{E}}^{*}_{-1} = \mathbf{I} - J^{1/3} \mathbf{U}^{-1} $ (Seth-Hill element of order -1) 
with a purely volumetric 
function $S_A(J)$. The   fourth order tensor $\bsym{\mathsf{A}}$,
specifies the shear properties at the rest state $J=1$. 
The choice of $\Phi(J)$ can be taken from formulations for equation of state, 
and then $S_A(J)$ can be associated in a similar way to the treatment of the shear modulus above.

For materials such as MgO whose anisotropy varies strongly with increasing pressure 
(Karki et al., 1997) we need to add an additional term to the strain energy, e.g., 
\begin{equation}
 S_B(J) \left[  \mathbf{E}^{*}_{-2} : \bsym{\mathsf{B}}: \mathbf{E}^{*}_{-2}  \right],
\label{e.29}
\end{equation}
in terms of the equivoluminal Eulerian strain 
${\mathbf{E}}^{*}_{-2} = \mathbf{I} - J^{2/3} \mathbf{U}^{-2} $ (Seth-Hill element of order -2) .
The function $S_B(J) = 0$ at $J=1$, and can be tuned to represent the variations in
anisotropy with  pressure.

\section{Conclusion}
We have shown how it is possible to develop formulations of nonlinear elasticity that can 
accommodate large shear and high compression, by the introduction of a shear function
as a function of volume modulating a deviatoric term.  For Earth materials, the functional 
dependence of the shear properties can be guided by the semi-empirical linear relation 
between shear modulus, bulk modulus and pressure.

For anisotropy a similar development can be made with functions of volume combined with 
strain energies depending on the the equivoluminal  components of the Seth-Hill family of 
strain tensors. 
The flexibility of the development provides a means of representing a wide range of isotropic and 
anisotropic scenarios suitbale for conditions in the Earth's interior.

The formulation developed in this work has been oriented toward situations with moderate to strong 
compression, but could also be used for strong expansion with a switch in the style
of strain measures employed. For compression, the deviatoric component is best represented 
using measures depending on strain exponent $q<0$, but in tension $q>0$ is to be preferred 
(Beex, 2019).

\appendix
\section*{Appendix}
\renewcommand\theequation{\mbox{\normalsize A.\arabic{equation}}}

The normalised stretch tensor $\mathbf{U}^{*}$ can be written in terms of its eigenvalues, 
the normalised stretches $\lambda_i^{*} = J^{-1/3}\lambda_i $, and their 
associated orthogonal eigenvectors  
$\mathbf{n}_i$ as:
\begin{equation}
\mathbf{U}^{*} = \sum_{i=1}^{3} \lambda_i^{*} \mathbf{n}_i \mathbf{n}_i  ,
\label{e.a1}
\end{equation}
in terms of the dyadic product of the eigenvectors.
The projection tensors $\bsym{\mathsf{P}}_q$ introduced in (\ref{e.24}) depend on 
the evolution of strain (Miehe \& Lambrecht, 2001; Beex, 2019) and can also be written 
in terms of the eigen-quantities:
\begin{equation}
\bsym{\mathsf{P}}_q = 2 \frac{\partial \mathbf{E}^{*}_q}{\partial \mathbf{E}} = 
\sum_{i=1}^{3} 
d^{\{q\}}_i \mathbf{n}_i \mathbf{n}_i \mathbf{n}_i \mathbf{n}_i 
+
\sum_{r=1}^{3}  \sum_{j \ne i}^{3} 
\vartheta^{\{q\}}_{ij} \big( \mathbf{n}_i \mathbf{n}_j \mathbf{n}_i \mathbf{n}_j 
+ \mathbf{n}_i \mathbf{n}_j \mathbf{n}_j \mathbf{n}_i  \big) .
\label{e.a2}
\end{equation}
The coefficients $d_i$ depend on the stretches and the order of the strain element $q$
\begin{equation}
d^{\{q\}}_i = \lambda_i^{*(q-2)} .
\label{e.a3}
\end{equation}
For three distinct stretches,
\begin{equation}
\vartheta^{\{q\}}_{ij} = \frac{1}{q}\frac{\lambda_i^{*q}- \lambda_j^{*q}}
{\lambda_i^{*2} - \lambda_j^{*2}} .
\label{e.a4}
\end{equation}
When two stretches are equal $\lambda^{*}_a = \lambda^{*}_b \ne \lambda^{*}_c$,
\begin{equation}
\vartheta^{\{q\}}_{ab} = \textstyle{\frac{1}{2}}\displaystyle d_a .
\label{e.a5}
\end{equation}
For the hydrostatic case, $\lambda^{*}_1 = \lambda^{*}_2 =\lambda^{*}_3 = 1$,
the coefficient $\vartheta^{\{q\}}_{ij} = \frac{1}{2}$.

Second derivative projection operators can be defined in a similar way in terms of the stretches 
and their associated eigenvectors, but now involve sixth-order tensors 
(Miehe \& Lambrecht, 2001; Beex, 2019).


\begin{references}
\hyphenchar\font = -1
\sloppy

\reference
Beex L.A.A. 2019.
Fusing the Seth–Hill strain tensors to fit compressible elastic material responses in the nonlinear regime.
\textit{Int. J. Mech. Sci.} \textbf{163} 105072

\reference
Burakovsky L., Preston D.L., Wang  Y., 2004.
Cold shear modulus and Gr\"uneisen parameter at all densities,
\textit{Solid State Commun.} \textbf{132}  151--156.

\reference
Destrade M., Odgen R.W., 2013.
On stress-dependent elastic moduli and wave speeds,
\textit{IMA J. Applied Mathematics}, \textbf{78}, 965--977.

\reference
Gr\'{e}aux S., Irifune T., Higo Y., Tange Y., Arimoto T., Liu Z., Yamada A., 2019.
Sound velocity of CaSiO$_3$ perovskite suggests the presence of basaltic crust in the 
Earth’s lower mantle. 
\textit{Nature} \textbf{565} 218-221. 

\reference
Hill R., 1968. 
On constitutive inequalities for simple materials—I. 
\textit{J. Mech. Phys. Solids} \textbf{16} 229--242.

\reference
Jackson I., Niesler H., 1982. 
The elasticity of periclase to 3GPa and some geophysical
implications. In: Akimoto, S., Manghnani, M.H. (Eds.), \textit{High-pressure Research in
Geophysics}, pp. 93–113, Centre of Academic Publications, Japan.

\reference
Karki B.B., Stixrude L., Clark S.J., Warren M.C., Ackland G.J., Crain J., 1997.
Structure and elasticity of MgO at high pressure.
American Mineralogist, \textbf{82}, 51–-60.

\reference
Kennett B.L.N., 2017.
Towards constitutive equations for the deep Earth.
\textit{Phys. Earth Planet. Inter.} \textbf{270}, 40--45.

\reference
Latorre M., Mont\'ans F.J., 2017.
WYPIWYG hyperelasticity without inversion formula: application to passive ventricular myocardium.
\textit{Comput. Struct.} \textbf{185} 47–-58.

\reference
Miehe C., Lambrecht M., 2001.
Algorithms for computation of stresses and elasticity moduli in terms of Seth-Hill’s family 
of generalized strain tensors. 
\textit{Commun. Numer. Methods Eng.} \textbf{17}, 337--353.

\reference
Mihai L.A., Goriely A., 2017. 
How to characterize a nonlinear elastic material?
 A review on nonlinear constitutive parameters in isotropic finite elasticity. 
\textit{Proc. R. Soc. A} \textbf{473} 20170607.

\reference
Seth B.R., 1964. 
Generalized strain measure with application to physical problems. In: 
\textit{Second-Order Effects in Elasticity, Plasticity and Fluid Dynamics}, 
Reiner M, Abir D (eds). Pergamon Press: Oxford,  162--172.

\reference
Sinogeikin S.V., Bass J.D., 2000. 
Single-crystal elasticity of pyrope and MgO to 20 GPa
by Brillouin scattering in the diamond cell.
\textit{Phys. Earth Planet. Int.} \textbf{120} 43–-62.

\reference
Spencer A.J.M., 1980.
\textit{Continuum Mechanics}, Longman.

\reference
Stacey F.D., Davis P.M., 2004.
High pressure equations of state with applications to the lower mantle and core.
\textit{Phys. Earth Planet. Inter.} \textbf{142} 137--184. 

\reference
Stixrude L., Lithgow-Bertelloni C., 2005.
Thermodynamics of mantle minerals - I. Physical Properties.
\textit{Geophys. J. Int.}, \textbf{162}, 610--632.

\reference
Zha C.-S., Mao H.-K., Hemley R.J., 2000. 
Elasticity of MgO and a primary pressure scale to 55 GPa. 
\textit{PNAS} \textbf{97} 13494–-13499.
\end{references}
\end{document}